\documentclass[11pt]{emulateapj}

\newfont{\myfont}{cmmib10}

\submitted{Ap.J. Lett. accepted}

\shorttitle{Faraday Rotation towards Sgr A*}
\shortauthors{Macquart et al.}

\begin{document}

\title{The Rotation Measure and 3.5\,mm Polarization of Sgr A*}

\author{Jean-Pierre Macquart\altaffilmark{1}, Geoffrey C. Bower\altaffilmark{2}, Melvyn C.H. Wright\altaffilmark{2}, Donald C. Backer\altaffilmark{2}, Heino Falcke\altaffilmark{3}}

\altaffiltext{1}{NRAO, P.O. Box 0, Socorro NM 
87801, U.S.A. {\it jmacquar@nrao.edu} (Jansky Fellow)}
\altaffiltext{2}{Astronomy Department and Radio Astronomy Laboratory, University of California, Berkeley, Berkeley, CA 94720, U.S.A.}
\altaffiltext{3}{ASTRON, Postbus 2, 7990 AA Dwingeloo, The Netherlands and Department of Astrophysics, Radboud Universiteit Nijmegen, Postbus 9010, 6500 GL Nijmegen, The Netherlands}
\begin{abstract}
We report the detection of variable linear polarization from Sgr A* at  a wavelength of $3.5\,$mm, the longest wavelength yet at which a detection has been made.  The mean polarization is $2.1 \pm 0.1$\% at a position angle of $16 \pm 2^\circ$ with rms scatters of 0.4\% and 9$^\circ$ over the five epochs.  We also detect polarization variability on a timescale of days.   Combined with previous detections over the range 150--400\,GHz (750--2000\,$\mu$m), the average polarization position angles are all found to be consistent with a rotation measure of $-4.4 \pm 0.3 \times 10^5\,$rad\,m$^{-2}$.  This implies that the Faraday rotation occurs external to the polarized source at all wavelengths.  This implies an accretion rate $\sim 0.2 - 4 \times 10^{-8}\, {\rm M}_\odot \,$yr$^{-1}$ for the accretion density profiles expected of ADAF, jet and CDAF models and assuming that the region at which electrons in the accretion flow become relativistic is within $10\,R_{\rm S}$.  The inferred accretion rate is inconsistent with ADAF/Bondi accretion. The stability of the mean polarization position angle between disparate polarization observations over the frequency range limits fluctuations in the accretion rate to less than $5$\%.  The flat frequency dependence of the inter-day polarization position angle variations also makes them difficult to attribute to rotation measure fluctuations, and suggests that both the magnitude and position angle variations are intrinsic to the emission.
\end{abstract}


\section{Introduction} \label{Introduction}

Linear polarization can be an important diagnostic of relativistic jets and accretion flows associated with black hole systems.  In the case of the massive black hole in the Galactic Center, Sgr A*, the properties of its mm-wavelength linear polarization probes the accretion environment on scales inaccessible with other techniques (Bower et al.\,1999a,b; Aitken et al.\,2001; Bower et al.\,2001; Bower et al.\,2003; Marrone et al.\,2006).  The apparent absence of linear polarization at wavelengths exceeding 2.7\,mm and sharp rise in polarization fraction at shorter wavelengths sets an upper limit to the rotation measure (RM).  This limits the mass accretion rate to $\sim 10^{-7} {\rm M}_\sun {\rm\ yr^{-1}}$ at distances of $10 - 1000$ Schwarzschild radii from the black hole, which eliminates certain classes of accretion flow (Quataert \& Gruzinov 2000; Agol 2000), but is consistent with CDAF and jet models (e.g. Falcke, Mannheim and Biermann 1993).  The RM measures the accretion rate by serving as a proxy for the electron column density once coupled with assumptions about the magnetic field.  Equipartition between kinetic, magnetic and gravitational energy is often assumed to relate the electron density and magnetic field (e.g. Bower et al. 1999a; Melia \& Falcke 2001; Marrone et al. 2006).

The discovery of variations in both the polarization angle (Bower et al.\,2005) and fraction (Marrone et al.\,2006) suggests two potential sources of variability.  The variations may be intrinsic and, therefore, offer evidence on the nature and structure of the emission region on a scale of $\sim 10 R_s$.  Changes in the RM along the line of sight may also induce external polarization variability. Structure in the accretion region on all scales within the Bondi radius can contribute to RM variations.  There is thus strong motivation to accurately characterize the RM, its fluctuations, and the intrinsic variability of the polarized source.

In \S2 we present the detection of linear polarization of Sgr A* at 3.5\,mm, showing that it varies on short timescales as well as relative to historical non-detections.  Combining this in \S3 with other data, our detection yields the best constraints so far on the RM.   We discuss the nature of the accretion flow and on the intrinsic source properties in \S4.  

\section{Observations} \label{Observations}
Sgr A* was observed with the BIMA array at 3.5 mm between 28 March and 01 April 2004 UT.  All five observations were obtained in identical LST time ranges that corresponded roughly
to 11 to 16 UT.  Observations were obtained in a polarimetric mode that produced Stokes $I$, $Q$, $U$ and $V$ images, with the two frequency bands centered on 82.76 and 86.31\,GHz, each with 800 MHz bandwidth.   Data were calibrated assuming that the flux density of J1733-130 was steady at its mean value of 2.7\,Jy.
J1733-130 was observed for 5 minutes every 30 minutes.  Observations were obtained in the B configuration giving a resolution of $8.6 \times 2.6\arcsec$ in position angle (p.a.) $7^\circ$.  The data were filtered to remove spacings shorter than 20\,k$\lambda$, similar to the processing of previous Sgr A* polarization experiments.   This $u$--$v$ distance cutoff gives a total flux estimate accurate to better than 10\% from 8.4 to 86 GHz (Bower et al.\,2001).  A priori amplitude calibration using system temperature and default gain information was applied to J1733-130.  Both sources were phase self-calibrated.  Polarization leakage terms from observations of 3C279 on 11 November 2003 were applied.  

Numerous sources of error contribute to the accuracy with which we can measure the polarization fraction and p.a..  In addition to statistical errors that include noise from the sky and the telescope, the leakage of polarization from one polarization handedness to the other is an important effect.  This converts a fraction of the total intensity into a false polarization signal.   Leakage calibration uses observations of a bright calibrator source and simultaneously solves for the polarization of the source and the leakage terms appropriate for individual antennas.  Imperfect correction leaves a residual false polarization signal and introduces errors that exist in the calibrator observation.  BIMA leakage solutions at 1.3\,mm were found to be stable over periods of months to years; variability in the solutions were due to changes in receiver orientation (Bower, Wright \& Forster, 2002). Typical variations in antenna leakage terms were $\sim 1.5\%$ over 2 years, indicating a $\sim 0.5\%$ variation in the average leakage solution for the 9 BIMA antennas.  Solutions at 3.5\,mm are expected to have the same characteristics although this has not been studied in depth.  In addition to variations in the leakage error, statistical errors in the calibrator data introduce errors.  A comparison of  actual measured p.a.s of 
3C 279 at 3.5\,mm and 1.3\,mm from BIMA with measured p.a.s at 1.3\,cm and 0.7\,cm from the VLA indicate excursions of 10$^\circ$-- 20$^\circ$ (Bower et al.\,2002).  These deviations exceed those expected from leakage corrections and statistical errors.  They are possibly real source effects but may also be uncorrected instrumental errors.
   
The average flux densities of Sgr A* and linear polarization E-vectors are listed in Table \ref{FluxTable}.  
In the polarization fraction and p.a. errors we include the effect of a 0.5\% polarization leakage error.  This error propagation is accurate in the case of the high SNR detections in the LSB but presents a lower limit to the errors in the low SNR detections in the USB.  Inter-day flux density variations are detected at a significant level, with a reduced-$\chi^2$ of 4.9 for the hypothesis of constant flux density for Sgr A* for 4 degrees of freedom.  The mean flux density is $1.93 \pm 0.07\,$Jy and the rms variability amplitude is $0.18\,$Jy.  

Polarization was detected in all five epochs, making this the first detection of polarization at 3.5\,mm, the longest wavelength at which polarization has been detected in Sgr A*.  The mean polarization
fraction is $2.1 \pm 0.1$\% and its mean p.a.~is $16 \pm 2^{\circ}$.  The polarization fraction and p.a.~vary from day to day.  The detection of polarization is more robust in the LSB (82.8 GHz) than in the USB (86.3 GHz).  The origin of this difference is unclear.  It may be due to changes in the leakage corrections and/or poor phase stability at the higher frequency.  We have previously seen data from BIMA in which polarization results in one sideband were more reliable than in the other 
(Bower et al.\,2001).  The bandwidth used in these observations is sufficiently small that the linear polarization is not depolarized by rotation of the polarization vector across the observing band, as discussed below.  


\begin{table}
\begin{tabular}{cccccccc}
\tableline
UT Day (2004)      &       $I$\,(Jy)  &   $I_{\rm err}$\,(Jy)  & $Q$\,(mJy)   &    $U$\,(mJy)    &   $\sigma$\,(mJy)   & P (mJy) & $\chi$ ($^\circ$)  \\
LSB \qquad \qquad \qquad & \null & \null & \null & \null & \null & \null & \null  \\
\tableline
88.5 & 1.94 & 0.11 & 20 & 34  &  5 & $39 \pm 5$   & $ 29 \pm 3$  \\
89.5 & 1.96 & 0.20 & 31 &  9   & 4  & $32 \pm  4$  &  $8 \pm 3$  \\
90.5 & 1.71 & 0.20 & 29 & 16  & 8  & $33 \pm  8$  & $14 \pm 6$  \\
91.5 & 1.69 & 0.20 & 29 & 18  & 8  & $34 \pm 8$   & $15 \pm 6$   \\
92.5 & 2.12 & 0.14 & 55 & 21  & 4  & $58  \pm 4$  & $10.0 \pm 1.0$ \\
USB \qquad \qquad \quad & \null & \null & \null & \null & \null & \null & \null  \\
\tableline     
88.5 & 1.98 & 0.12 & -10 &   11 &  5 & $14 \pm 5$ &  $66 \pm 9$ \\
89.5 & 1.97 & 0.18 &   7   & -15 &  4  & $16 \pm 4$ & $-32 \pm  6$ \\
90.5 & 1.68 & 0.19 &   3   &     3 &  7  & $4 \pm  7$  & $22 \pm  47$ \\
91.5 & 1.76 & 0.16 &   9   &     6 &  7 & $10 \pm 7$  & $16 \pm 18$ \\
92.5 & 2.25 & 0.09  & 30  &  19 &  4 & $35 \pm 4$   & $16 \pm 3$ \\
\tableline
 \end{tabular}
 \caption{The daily average flux densities of Sgr A*.  $I_{\rm err}$ is determined from 
the flux density scatter determined on short timescales for each day.  $\sigma$ is the statistical error for $Q$ and $U$; errors in the polarization fraction and p.a.~include the effects of 0.5\% leakage error.} \label{FluxTable}
\end{table}

\begin{table}
\begin{tabular}{cccccc}
\tableline
 $\nu$ (GHz)  &	$P_l$(\%) & $\sigma_{P}$ (\%)   &    $\chi$ ($^\circ$)  & $\sigma_\chi$ ($^\circ$) & Measurement epochs  \\
\tableline
 82.8  	& 2.1  & 0.4 	&  15.2 & 8.2  & 5 \\
 86.3    & 0.8  & 0.5          &  18 & 35  & 5 \\
150$^a$ 		& 12 & $ \null_{-4}^{+9}$  	&  83 & 3 & 1\\
222$^a$ 		&  11 & $\null_{-2}^{+3}$ 	&  88  & 3  & 1\\
216$^{b}$ 	&  10 & 1 	& 115 & 13 & 2 \\	
230$^{b}$	& 9 & 3	& 117 & 24  & 7 \\	
340$^{c}$	&  6.1 & 2.0  	& 145 & 9 & 6 \\
350$^a$		&  13 & $\null_{-4}^{+10}$ 	&  161 & 3 &  1\\
400$^a$  		&  22 & $\null_{-9}^{+25}$  	&  169 & 3  & 1\\
\tableline
\end{tabular}
 \caption{Mean polarization fractions and p.a.s from measurements of $^a$Aitken et al. 2001,$^b$Bower et al. 2003 \& 2005, and $^c$Marrone et al. 2006.  The quantities $\sigma_P$ and $\sigma_\chi$ denote the rms variation in the polarization fractions and p.a.s respectively except in the case of the single-epoch measurements, where they denote the estimated error of the measurement.} \label{PolnTable}
\end{table}

\section{The Rotation Measure} \label{RMSect}


Table \ref{PolnTable} lists the mean polarization p.a.s for all previous detections of linear polarization in Sgr A*.  Due to systematic errors from polarization leakage calibration, the poor quality of upper sideband data at 3.5\,mm, and the relatively small spacing between the two sidebands, we are unable to compute a meaningful RM estimate from the 3.5\,mm data alone.  The combined average polarization data are consistent with a single RM of $-4.4 \pm 0.3 \times 10^5\,$rad\,m$^{-2}$ and an intrinsic polarization p.a.~$\chi_0=168 \pm 8^\circ$.  Fig.\,1 plots the best RM fit to the mean polarization p.a. data.  The best fit, with a reduced-$\chi^2$ of 29 computed using the errors quoted in Table \ref{PolnTable}, is obtained if the 85\,GHz p.a. is rotated by $-180^\circ$.  A fit to the unwrapped 85\,GHz p.a. yields a reduced-$\chi^2$ of 54 and ${\rm RM}=-2.0 \pm 0.4 \times 10^5\,$rad\,m$^{-2}$, while an extra $-180^\circ$ wrap yields ${\rm RM}=-6.9 \pm 0.4 \times 10^5\,$rad\,m$^{-2}$ with an associated reduced-$\chi^2$ of 64.

The reduced-$\chi^2$ for the best fit is unacceptably high because of the unrealistically small errors associated with the Aitken et al.\,(2001) observations.  These single-epoch low-resolution measurements predate the identification of polarization variability in Sgr A*.   We therefore fit only to interferometric observations spanning multiple epochs.  The resulting fit properties depend on whether one regards measurements from the two sidebands at $\nu> 216\,$GHz, where available, as being mutually independent.  If so, the reduced-$\chi^2$ for a fit including the two -180$^\circ$-wrapped 3.5\,mm points is 0.9, but is 3.6 for the unwrapped 3.5\,mm polarization pas.   If not, the two $\chi^2$ values are 0.8 and 2.1 respectively.  In both cases, the solution involving the wrapped 3.5\,mm points is  preferred.  We regard these cases as bounding the true significance of the RM.  For the case of independent sideband measurements, the associated best fit is $\chi_0 = 163 \pm 2^\circ$ and ${\rm RM}= -4.38 \pm 0.06 \times 10^5\,$rad\,m$^{-2}$.  

We adopt the RM $=-4.4 \pm 0.3 \times 10^5\,$rad\,m$^{-2}$ derived from
all of the multi-epoch data.  Given the variability of the p.a.~and the uncertain 
systematics of different measurements, we consider the use of all data to provide the most 
conservative estimate of the RM.

\begin{figure}[h]
 \includegraphics[angle=0,scale=0.6]{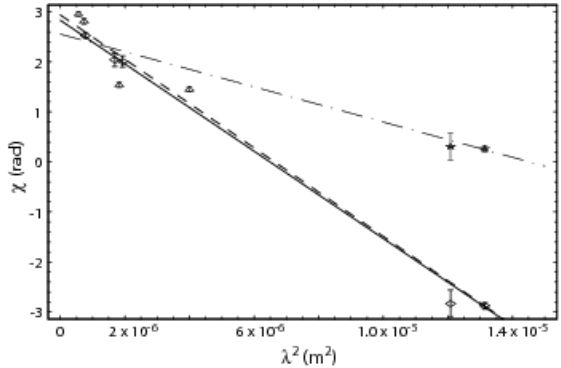}
\caption{The mean polarization p.a.~of Sgr A* as a function of wavelength.  The 85\,GHz points have been derotated by $-180^\circ$.  Diamonds denote measurements for which polarization variability is detected, and their error bars denote the standard error of the mean of the p.a.~variations.  Triangles denote single epoch measurements only.  The dashed line shows the best fit to all data, the solid line to all points excluding the Aitken et al. (2001) data, and the dot-dashed line to the unwrapped 85\,GHz points, denoted by stars.} \label{Figlabel}
\end{figure}

\begin{figure}[h]
 \includegraphics[angle=0,scale=0.7]{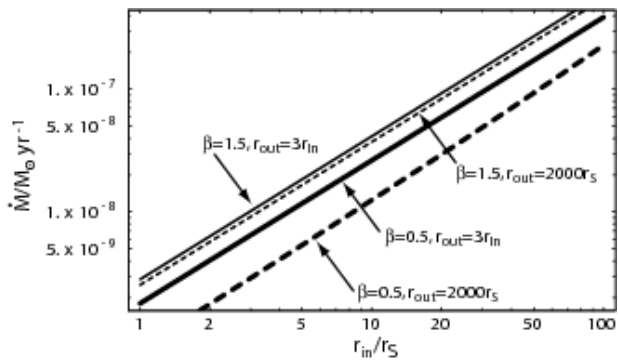}
\caption{The accretion rate implied by our measurement of the RM for various accretion models using $M_{\rm bh} = 2.6 \times 10^6\,$M$_\odot$ (Ghez et al.\,1998).  Two choices for $r_{\rm out}$ are shown: one in which the outer scale is large (effectively infinite) and another in which it is only three times larger than the inner cutoff radius.} \label{FigMdotlabel}
\end{figure}

\section{Discussion} \label{Discussion}

The accretion rate implied by this RM depends on density and magnetic profiles assumed.  Following the prescription outlined by Marrone et al.\,(2006),\footnote{Note the typographical error in eq.\,(9) of Marrone et al.\,(2006) in which one should have RM$ \propto r_{\rm in}^{-7/4}$.} we write $n \propto r^{-\beta}$, for  $r_{\rm in} < r <r_{\rm out}$ and assume equipartition between magnetic, kinetic and gravitational energy.  The radius $r_{\rm in}$ designates the point at which the flow becomes relativistic, and so ceases to contribute to the RM because its contribution to Faraday rotation is decreased by a factor $\gamma^{2}/\log \gamma$.  The outer scale is a proxy for the coherence length of the magnetic field fluctuations.  The estimated accretion rate for various ADAF ($\beta=3/2$) and CDAF ($\beta=1/2$) models is shown in Figure \ref{FigMdotlabel} as a function of $r_{\rm in}$.  
The accretion rate is in the range $0.2 - 4 \times 10^{-8}\,$M$_\odot$\,yr$^{-1}$ for $2 \, R_{\rm S} < r_{\rm in} < 10 \,R_{\rm S}$, $r_{\rm out} > 3 \, r_{\rm in}$ and $1/2<\beta < 3/2$.

The fact that a single RM accounts for the frequency dependence of all mean polarization p.a.s implies that the Faraday rotation occurs external to the polarized source.  Internal rotation would cause the RM to vary as a function of frequency.  The emission is optically thick at all frequencies at which polarization is detected (Falcke et al.\,1998; Zhao et al.\,2003, Yuan et al.\,2003), so if any internal Faraday rotation did occur, the diminuation of opacity effects with frequency, which increases the depth down to which one observes emission, would cause a corresponding increase in the Faraday rotation path length.  


%
An interpretation of the p.a.~jitter observed at 85\,GHz in terms of RM variability would imply rms deviations of $1.2 \times 10^4\,$rad\,m$^{-2}$, far smaller than the $\sim 2 \times 10^5\,$rad\,m$^{-2}$ rms deviations implied by the 340\,GHz Marrone et al.\,(2006) fluctuations interpreted similarly.
However, the absence of a clear frequency dependence in the inter-day jitter makes it difficult to ascribe to RM fluctuations, and hence variations in the accretion rate.  Table 2 shows the rms p.a. deviations for the multi-epoch observations at 85, 216, 230 and 340\,GHz.  These are inconsistent with the $\nu^{-2}$ dependence expected of RM fluctuations from a magnetoionic medium external to the source, suggesting instead that the jitter reflects changes in the intrinsic source polarization p.a..  Nonetheless, the fact that the fluctuations at these frequencies were not observed simultaneously, coupled with the short time span of the 85\,GHz observations compared to the $>2\,$month -- albeit sporadic -- sampling of the 230 and 340\,GHz measurements, still admits the possibility that the p.a.~dispersion observed at 85\,GHz is unrepresentative of its long-term average.  This appears unlikely.  
If the $\approx 20^\circ$ rms p.a. variations observed at 216--230\,GHz were associated with RM fluctuations we would expect $\approx 150^\circ$ fluctuations at 85\,GHz and would not expect to find $\chi \propto \lambda^2$ over a set of disparate measurements.  
Moreover, the presence of intrinsic p.a. changes is unsurprising given that the polarization fraction is also intrinsically variable, as discussed below.

\begin{figure}[h]
\vskip 5mm
 \includegraphics[angle=0,scale=0.9]{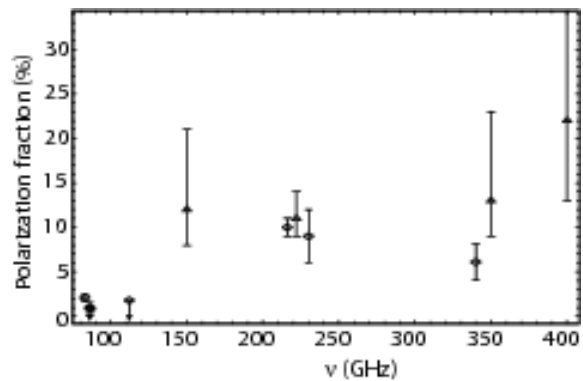}
\caption{The mean polarization fraction of Sgr A*.  The error bars plotted are those of Table \ref{PolnTable}.  Previous 86 and 112\,GHz non-detections are marked with arrows.  The error bars in the Aitken et al.\,(2001) measurements, marked with triangles, reflect uncertainty in the contribution from dust emission rather than variability associated with the source.} \label{FigPolnlabel}
\end{figure}

The lack of p.a.~jitter attributable to RM fluctuations can be interpreted in terms of an upper limit in accretion rate fluctuations.  The absence of clearly identifiable inter-day RM fluctuations suggests it is uniform on inter-day timescales. The consistency of the observations over a large number of disjoint epochs and frequencies with a single RM further suggests that the underlying accretion rate is constant on the timescale over which the observed polarization pas were averaged.  
The uncertainty in our RM fit places an upper bound on the accretion rate variations using ${\rm RM} \propto \dot{M}^{3/2}$, valid under the assumption of equipartition between magnetic, kinetic and gravitational energy (e.g. Marrone et al.\,2006).  The 7\% uncertainty in the RM implies $\dot{M}$ fluctuations less than $5$\%.  The limit on $\dot{M}$ variations is largely consistent with the limits imposed by source flux density variations.  In the jet model the flux density scales as $\dot{M}^{17/12}$ (Falcke et al.\,1993).  The standard deviation of the 3.5\,mm fluxes is 10\%, comparable to the errors on the individual measurements, which imposes an upper limit of 14\% on $\dot{M}$ fluctuations.
Marrone et al.\,(2006) detect 10\% rms intensity variations at 340\,GHz.  The five intensity measurements from Bower et al.\,(2005) at 216--230\,GHz exhibit 29\% modulations, implying $\dot{M}$ fluctuations of 40\%.  Variability in polarization fraction is detected by all multi-epoch observations (Fig.\,\ref{FigPolnlabel}), at 85, 230 and 340\,GHz, and presumably occurs at 112\,GHz, where it was not detected at the 1.8\% (1-$\sigma$) level by a previous search (Bower et al. 2001).  A previous limit of 1\% linear polarization at 86\,GHz (Bower et al.\,1999b) demonstrates that it also varies on long timescales.

Since the Faraday screen is external to the source the polarization amplitude fluctuations must be intrinsic.  Instrumental depolarization effects are too low to explain the variability: bandwidth depolarization is only important for RMs greater than $2.7 \times 10^7\,$rad\,m$^{-2}$ at 85\,GHz, while beam depolarization is similarly improbable given the sub-mas size of  Sgr A* at mm wavelengths (Bower et al. 2004; Shen et al. 2005).  Spatial variation in the RM across the transverse extent of the source could depolarize the emission, but at even 85\,GHz this requires RM fluctuations $\delta {\rm RM} \ga 7 \times 10^5\,$rad\,m$^{-2}$ (Quataert \& Gruzinov 2000).

Both the high variability of the mm and sub-mm emission  (Zhao et al.\,2004; Wright \& Backer 1993; Tsuboi et al.\,1999) and the linearly polarized emission of Sgr A* are possibly associated with its excess sub-mm emission (Serabyn et al.\,1997; Falcke et al.\,1998; Melia et al.\,2001).  ADAF models that fit the cm to-sub-mm spectrum include at least two distinct populations of radiating particles (Yuan et al.\,2003), with the second important at $\nu \ga 100\,$GHz in order to explain the sub-mm bump.

The model of Yuan et al.\,(2003), in which the sub-mm emission is dominated by thermal electrons, overpredicts the polarization fraction at 85\,GHz (cf. Fig.\,\ref{FigPolnlabel}).  In this model the degree of linear polarization ranges from 32\% at 85\,GHz to 70\% at 400\,GHz assuming a uniform magnetic field and that Faraday depolarization intrinsic to the source is unimportant (see their Fig.\,3b). In the context of this model the ratio of the predicted to observed polarization levels can only be attributed to magnetic field inhomogeneity intrinsic to the source.  This ratio is a factor of 3 higher at 85\,GHz relative to the fraction in the range $150-400\,$GHz, over which it is constant within the errors.  It is hard to account for such an increase in magnetic field inhomogeneity at 85\,GHz, particularly if the source size only scales $\propto \nu^{-1}$, as expected if its emission is self-absorbed.  On the other hand, including synchrotron self-absorption effects, Goldston, Quataert \& Igumenshchev (2005) show that the polarization fraction is expected to increase by a factor of three over the range $85-200\,$GHz.
The quasi-spherical accretion polarization model of Melia, Liu \& Coker (2000) and the two-component model of Agol (2000) predict a 90$^\circ$ p.a.~flip at $\sim 280\,$GHz which is at variance with the measured p.a.s at 150, 230 and 340\,GHz.

We have reported here the detection of linear polarization in Sgr A* at 3.5\,mm.  This enables us to calculate the rotation measure and set a limit on the accretion rate.  The lack of frequency dependence for position angle fluctuations indicates that
they are intrinsic to the source.  Our result favors RIAF/CDAF accretion models, with a shallow density distribution, over ADAF and Bondi-Hoyle accretion flows, which have a steep profile and are more likely to produce rapid RM variations.
Future wide bandwidth simultaneous observations with CARMA and the SMA will fully characterize intrinsic and extrinsic changes in the polarization properties of Sgr A* and allow us to investigate the accretion environments of other nearby low luminosity AGN, such as M81* (Brunthaler, Bower \& Falcke 2006).


\begin{thebibliography}{}
\bibitem[Agol(2000)]{Ag00} Agol, E. 2000, \apj, 538, L121
\bibitem[Aitken et al.(2001)]{Aik01} Aitken, D.K., Greaves, J., Chrysostomou, A., Jenness, T.,  Holland, W., Hough, J.H., Pierce-Price, D. \& Richer, J. 2001, \apj, 534, L176
\bibitem[Bower et al.(1999a)]{Bow99a} Bower, G.C., Backer, D.C., Zhao, J.-H., Goss, W.M. \& Falcke, H. 1999a, \apj, 521, 582
\bibitem[Bower et al.(2004)]{Bow04} Bower, G.C., Falcke, H., Herrnstein, R.M., Zhao, J.-H., Goss, W. M. \& Backer, D.C. 2004, Science, 304, 704
\bibitem[Bower et al.(2005)]{Bow05} Bower, G.C., Falcke, H., Wright, M.C.H. \& Backer, D.C. 2005, \apj, 618, L29 
\bibitem[Bower et al.(1999b)]{Bow99b} Bower, G.C., Wright, M.C.H., Backer, D.C. \& Falcke, H. 1999b, \apj, 527, 851
\bibitem[Bower et al.(2001)]{Bow01} Bower, G.C., Wright, M.C.H., Falcke, H. \& Backer, D.C. 2001, \apj, 555, L103
\bibitem[Bower et al.(2003)]{Bow03} Bower, G.C., Wright, M.C.H., Falcke, H. \& Backer, D.C. 2003, \apj, 588, 331
\bibitem[Bower, Wright \& Forster (2002)]{BWF02} Bower, G.C., Wright, M.C.H. \& Forster, J.R. 2002, ``Polarization Stability of the BIMA Array at 1.3\,mm'', BIMA Memos \#89
\bibitem[Brunthaler et al.(2006)]{Br06} Brunthaler, A., Bower, G.C. \& Falcke, H. 2006, \aap, 451, 845\bibitem[Falcke, Mannheim \& Biermann (1993)]{FMB93} Falcke, H., Mannheim, K. \& Biermann, P.L. 1993, \aap, 278, L1
\bibitem[Falcke et al.(1998)]{Falcke98} Falcke, H., Goss, W.M., Matsuo, H., Teuben, P., Zhao, J.-H. \& Zylka, R. 1998, \apj, 499, 731
\bibitem[Ghez et al. 1998]{G98} Ghez, A.M., Klein, B.L., Morris, M. \& Becklin, E.E. 1998, \apj, 509, 678
\bibitem[Goldston et al.(2005)]{GQI05} Goldston, J.E., Quataert, E. \& Igumenshchev, I.V. 2005, \apj, 621, 785
\bibitem[Marrone et al.(2006)]{Mar06} Marrone, D.P., Moran, J.M., Zhao, J.-H. \& Rao, R. 2006, \apj, 640, 308
\bibitem[Melia \& Falcke (2001)]{MF01} Melia, F., Falcke, H. 2001, \araa, 39, 309
\bibitem[Melia, Liu \& Coker (2000)]{MLC00} Melia, F., Liu, S. \& Coker, R. 2000, \apj, 545, L117
\bibitem[Melia et al.(2001)]{Mel01} Melia, F., Liu, S. \& Coker, R. 2001, \apj, 553, 146 
\bibitem[Quataert \& Gruzinov (2000)]{QG00} Quataert, E. \& Gruzinov, A. 2000, \apj, 545, 842
\bibitem[Serabyn et al.(1997)]{Se97} Serabyn, E., Carlstrom, J., Lay, O., Lis, D.C., Hunter, T.R., Lacy, J.H. \& Hills, R.E. 1997, \apj, 490, L77 
\bibitem[Shen et al.(2005)]{Sh05} Shen, Z.-Q., Lo, K.Y., Liang, M.-C., Ho. P.T.P. \& Zhao, J.-H. 2005, \nat, 438, 62 
\bibitem[Tsuboi et al.(1999)]{T99} Tsuboi, M., Miyazaki, A. \& Tsutsumi, T. 1999, in The Central Parsecs of the Galaxy, ASP Conference Series, Vol. 186. eds. H. Falcke, A. Cotera, W.J. Duschl, F. Melia \& M.J. Rieke, 105
\bibitem[Wright \& Backer (1993)]{WB93} Wright, M.C.H. \& Backer, D.C. 1993, \apj, 417, 560
\bibitem[Yuan et al.(2003)]{Yu03} Yuan, F., Quataert, E. \& Narayan, R. 2003, \apj, 598, 301
\bibitem[Zhao et al. (2003)]{Zhao03} Zhao, J.-H., Young, K.H., Herrnstein, R.M., Ho, P.T.P., Tsutsumi, T., Lo, K.Y., Goss, W.M. \& Bower, G.C. 2003, \apj, 586, L29
\bibitem[Zhao et al.(2004)]{Zhao04} Zhao, J.-H., Herrnstein. R.M., Bower, G.C., Goss, W.M. \& Lui, S.M. 2004, \apj, 603, L85
\end{thebibliography}
\end{document}